\documentclass[pre,twocolumn,floatfix,nobibnotes]{revtex4}
\usepackage{graphicx,amsfonts,amssymb,amsmath, hyperref,multirow,array}
\usepackage[utf8]{inputenc}

\newlength{\ldag}
\begin{document}

\title{The effect of water on granular liquid flows: from debris to mud flows}

\author{Olivier Coquand}
\email{olivier.coquand@univ-perp.fr}
\affiliation{Laboratoire de Modélisation Pluridisciplinaire et Simulations, Université de Perpignan Via Domitia,
52 avenue Paul Alduy, F-88860 Perpignan, France}

\begin{abstract}
	In this work, we show how the rheology of granular suspensions can be related to the properties of the fluctuations of the
	velocity field inside the medium.
	In particular, effective Navier-Stokes equations in the different flow regimes are constructed and compared to an actual geophysical model
	that was so far purely phenomenological.
	Then, it is shown that a direct cascade of kinetic energy is present when the flow becomes turbulent, but with a scaling law that is
	quantitatively very different from that of usual Newtonian fluids.
\end{abstract}

\maketitle

\section{Introduction}

	Since the seminal work of the Groupe de Recherche sur les Milieux Divisés \cite{GDR04}, the laws of rheology of granular matter in dense
	flows have been established in a form that still constitutes today the standard in the field.
	More precisely, it has been successfully shown that granular fluids divide into two classes of materials \cite{Andreotti13}: the dilute granular
	gases that are mostly found in astrophysical systems, and a distinct state, called the \textit{granular liquid}, characterised by packing fractions
	bigger than 40\% and that obeys the flow rules exhibited in \cite{GDR04} with remarkable universality. Indeed, the flow rules are largely independent
	on the monodispersity of the granular packing, the form of its grains and their relative microscopic friction coefficient for example.

	This last property is crucial as a theory of the behavior of the granular liquid explaining the experimental results from first principles remained elusive for
	nearly 20 more years.
	To the best of our knowledge, the bases of such a theory were first laid down by Kranz et al. in \cite{Kranz18}, and then completed into a full consistent
	theory in successive studies \cite{Kranz20,Coquand20f,Coquand20g,Coquand21}.
	This model, called the Granular Integration Through Transients model, or GITT, (after the Integration Through Transients method built by Fuchs and Cates
	\cite{Fuchs02,Fuchs03,Fuchs09}) has shown a remarkable quantitative agreement with experimental data \cite{Coquand20f,Angelo23,Angelo25}.

	These studies confirmed the result first shown by De Giuili et al. \cite{DeGiuli15,Degiuli16} that the liquid state of granular matter -- note that we refer
	here to \textit{states} rather than \textit{phases} of matter because granular systems are always out of equilibrium -- does not extend up to the solid state;
	before the system effectively coalesces, it enters a regime controlled by the interparticle friction that represents only a few percents in terms of packing
	fraction, but has drastic consequences as it entirely sets the values of the critical exponents at the jamming transition for example.
	This is the regime referred to as \textit{quasi-static} in part of the literature.

	Going further, more recent studies have tried to establish a parallel between the fluctuations in granular flows and those observed in turbulent flows of
	Newtonian fluids in the form of a cascade picture of kinetic energy spectrum repartition in the Fourier space similar to that exhibited by Kolmogorov
	in the early 1940's \cite{Kolmogorov41a,Kolmogorov41b,Kolmogorov41c}. The point of such models is to better understand how energy is shared between the
	fluctuation modes at different scales and thus enter in greater detail into the mechanisms at the basis of the observed macroscopic rheology.
	Unfortunately, there are, to the best of our knowledge, no experimental study of granular turbulence, and the numerical studies give so far results
	that are a bit difficult to properly interpret: (i) in two dimensions, there have been several studies since the seminal paper of Radjai and Roux
	\cite{Radjai02}: \cite{Richefeu12,Combe15,Saitoh16a,Kharel17}, but they lead to different exponents for the cascade power laws, that are nonetheless
	not necessarily exclusive with each other given the difficulty of obtaining statistics of sufficiently good quality; (ii) in three dimensions there
	is to the best of our knowledge only one study, \cite{Saitoh16a}, but for the same reasons as in 2D, it is quite difficult to assess the quality of the
	predicted exponent. For all the above reasons, this topic can be considered to stand only at its early stage of study.

	In a recent work \cite{Coquand25}, we made a first attempt at building a theoretical model of fluctuations in a driven dry granular liquid.
	While qualitative arguments point out to a power law for the decay of the energy spectrum compatible with the numerical results \cite{Saitoh16a}, and
	thus different from the usual Kolmogorov exponent observed in Newtonian fluids, our study revealed that the symmetry breaking leading to this
	result is quite subtle, and more work is required for the model to yield solid conclusions.
	It is also noteworthy that the predicted exponent in that case, $3/2=1.5$, is numerically quite close to the Kolmogorov exponent $5/3\simeq1.67$,
	so that very good quality data is required to be able to distinguish one exponent from the other without ambiguity.

	The short review above only concerns dry granular flows however.
	In the case where granular particles are in suspension in a viscous liquid, which is typically the case when a dry soil becomes wetted by heavy rain
	for example, much less is known.
	For starters, even though attempts at founding an analogous law for the rheology of granular suspensions have been proposed -- the most convincingly
	connected to experiments being in our opinion \cite{Boyer11} -- there is nowadays still no consensus on this matter (see \cite{Tapia19} and references therein).
	An explanation of this situation has been proposed in \cite{Coquand20g}: the fact that a very simple yet efficient law of rheology has been found
	so easily in the dry case is an "accident" linked to the fact that most of the experiments are conducted in a particular regime of motion,
	called the \textit{Bagnold regime}.
	When the particular conditions giving the Bagnold equation are not met, the laws of rheology are modified, and one more dimensionless number
	is needed for a proper description of the rheology.
	In granular suspensions, the typical experimental setup is not in the Bagnold flow regime, so that the two dimensionless numbers are almost
	always needed \cite{Coquand20g}.
	Yet, if we agree to set aside the one dimensionless number constraint, the laws of rheology can be expressed in simple terms and in a common
	way in dry granular liquids and granular suspensions thanks to the GITT model \cite{Coquand20g,Coquand21}.
	However, in contrast to the dry case, the GITT model still misses an experimental validation in granular suspensions to solidly prove its validity.

	In the following article, we propose to explore the question of how the presence of a viscous liquid surrounding the granular particles influences
	the properties of its flow.
	This situation typically corresponds to the case of a soil that, by getting wetted, gets different flow properties, what can lead to catastrophic events
	like landslides and mud flows.
	Obviously, such events cannot be predicted exactly, but by giving a view on how the granular suspension responds to stress in contrast to its dry counterpart
	in a model based on first principles, we aim at helping improve the models that can predict how such events evolve through time.
	In particular, our study will yield two strong predictions that can be matched to concrete experiments: (i) first, we are going to show that the GITT model
	can help understand why the usual laws of dry granular liquids fail to describe some debris flow, and that its predictions are in agreement with
	phenomenological laws already used by the volcanology community; (ii) second, we are going to examine the cascade argument of von Weizsäcker \cite{Weizsacker48}
	to predict how fluctuations lead to the formation of a robust direct energy cascade (in three dimensions) with a power law exponent of 3, quantitatively
	very different from that known for Newtonian liquids, or presumed for dry granular liquids.

	This paper is organised as follows: the first section is a reminder of the most important properties of the rheology and flow of granular liquids, that
	can also be helpful as a dictionary to bridge the gap between the different communities interested in our problem.
	Then, we are going to show how the GITT predictions can be implemented in an effective Navier-Stokes equation.
	This formulation will be helpful to examine depth-averaged models used to calibrate and predict debris and pyroclastic flows.
	The next section is dedicated to the study of the fluctuations of the velocity field, which is connected to the cascade of energy argument.
	Finally, we conclude.

\section{Flowing properties of granular liquids}

	In this section, we briefly recall the main properties of granular flow rheology without entering into the complexity of the underlying models.
	For a complete derivation of the equations below, the interested reader is referred to the original paper \cite{Coquand20g}.

	\subsection{Rheology of dry granular liquids}

		The rheology of granular matter in the liquid state depends on two parameters: the packing fraction $\varphi$, and the shear rate $\dot\gamma$.
		The flow is governed by two equations: a dynamical equation (which takes the form of a Mori-Zwanzig equation in the GITT model) that we chose not
		to discuss explicitly here for sake of simplicity since it will not really influence our discussion, and the power conservation equation which
		takes the following form:
		\begin{equation}
		\label{eqE}
			\sigma\dot\gamma = \mathcal{P} + \rho \Gamma_d\,\omega_c\,T
		\end{equation}
		where $\sigma$ is the shear stress, thus the term on the left-hand side corresponds to shear heating, $\mathcal{P}$ is a generic term encompassing
		all possible sources of energy injection or dissipation other than shear heating and dissipation by collisions, $\rho$ is the mass density of the
		system, $\Gamma_d$ is a term quantifying the power dissipated in a single collision event, if $\varepsilon$ is the restitution coefficient of the
		granular particle, $\Gamma_d = (1 - \varepsilon^2)/3$ is generally a good estimate, $\omega_c$ is the collision frequency; its value can
		be estimated from the value computed from Chapman-Enskog models of granular gases $\omega_c = 24\varphi\chi d^{-1}\sqrt{T/\pi}$, $\chi$ being
		the value of the pair correlation function at contact (quantifying the probability that two particles collide at a given packing fraction)
		and $d$ the diameter of a granular particle, and finally, $T$ is the granular temperature,
		defined from the second moment of the velocity fluctuation distribution.
		As a consequence, it is a purely kinetic, rather than thermodynamic, temperature.

		Despite the many terms described above, Eq.~(\ref{eqE}) is a simple power budget equation.
		In most experiments on dry granular liquids, $\mathcal{P}=0$.
		For example, in a chute, or a rotating drum experiment, gravity enters the power balance equation as a source of shear, and collisions are the only source
		of dissipation.
		In such cases, where the only source of energy injection is shear heating and the only source of energy dissipation is collisions (that is, $\mathcal{P}=0$),
		Eq.~(\ref{eqE}) takes a particular form called the \textit{Bagnold equation}.

		Understanding this last point is key to understanding a lot of peculiarities of granular flows compared to other kind of fluids.
		Indeed, in a colloidal suspension for example, such an equilibrium never exists.
		Also, putting $\mathcal{P}=0$ in the power budget equation reduces the number of available degrees of freedom by one, with drastic consequences, as
		we are going to discuss below.

		Before entering into more details into the laws of rheology, let us also present the three time scales that govern its behavior: the macroscopic
		time scale $t_\gamma=1/\dot\gamma$ called the \textit{advection time scale}, the time scale separating two collision events $t_{ff}=d\sqrt{\rho/P}$,
		where $P$ is the pressure in the granular liquid, called the \textit{free fall time scale}, which can also be understood in a simpler
		way by the property $t_{ff}\propto1/\omega_c$ with a proportionality coefficient of order 1 \cite{Coquand20g}, and finally a time scale of the mesoscopic
		collective motion $t_\Gamma$ that can be built from the effective diffusion coefficient $D_{eff}$ of a granular particle in the fluid as
		$t_\Gamma = d^2/D_{eff}$.
		At low packing fractions, the collision events are rare enough for collective motions to be reduced to clusters of the order of one particle
		so that $t_\Gamma\simeq t_{ff}$, but as the medium becomes denser, this equation does not hold anymore and the two time scales decouple, until
		the so-called \textit{granular glass transition} is reached and $D_{eff}\rightarrow+\infty$, so that $t_\Gamma\rightarrow0$.
		In our GITT model, the granular glass transition takes place for $\varphi\simeq0.585$, but the quantitative predictions of this model for the
		values of limiting packing fractions can be quite different from the experimental ones and only the order of magnitude is important here
		(see \cite{Coquand20f} for a detailed discussion).

		Putting all of this together, the GITT model predicts three different flow regimes for granular liquids (in agreement with experiment):
		\begin{itemize}

			\item At low packing fractions and low shear rate, that is $t_\gamma\gg t_\Gamma\simeq t_{ff}$, the liquid is in a Newtonian regime.
				This regime is characterised by a constitutive relation of the type:
				\begin{equation}
					\sigma = \nu\,\dot\gamma
				\end{equation}
				where the constant $\nu$ is called the viscosity of the liquid.
				Hence, the granular liquid under these conditions behaves as an ordinary Newtonian liquid.

			\item At similar shear rates, but above the granular glass transition, the liquid enters the \textit{yielding regime}.
				In terms of time scales, because $D_{eff}$ diverges, the mesoscopic and microscopic time scales now decouple.
				However, the shear rate is still low at the scale of the collision events: $t_\gamma \gg t_{ff}$, $t_\Gamma=0$.
				In this regime,
				\begin{equation}
					\sigma \simeq \sigma_y
				\end{equation}
				where $\sigma_y$ is a material dependent constant.
				The behavior of the system is that of a yielding fluid: if it is submitted to a stress $\sigma<\sigma_y$, it is not put into motion,
				that is, the system does not flow.
				Note that below the granular glass transition, the viscosity coefficient is linked to the yield stress by the relation
				$\nu = t_\Gamma\,\sigma_y$.

			\item Finally, when the Bagnold equation holds, a new constraint is added on the system. Typically, from the above, we can deduce that
				$T\propto \big(\sigma\dot\gamma\big)^{2/3}$. Since $\sigma_y\propto T$, or equivalently below the granular glass transition
				$\nu\propto \sigma_y\propto T$, the relation between $\sigma$ and $\dot\gamma$ is modified.
				Rearranging the terms, we now get:
				\begin{equation}
					\sigma = B\,\dot\gamma^2
				\end{equation}
				where $B$ is called the Bagnold coefficient.

				A few comments are in order here.
				First, the GITT model has a smooth elastic limit ($\varepsilon\rightarrow1$), however since in this limit the collisions do not dissipate
				energy anymore, the Bagnold equation cannot hold and only the two first flowing regimes are found in elastic systems such as colloidal
				suspensions for example (this property is enforced by the fact that $\Gamma_d\rightarrow0$ when $\varepsilon\rightarrow1$).

				Second, as discussed in the introduction, typical experiments on granular liquids naturally occur in the Bagnold regime.
				But in this regime, either $t_\Gamma\simeq t_{ff}$ (below the granular glass transition), or $t_\Gamma=0$ (above the transition).
				Hence, the whole range of packing fractions can entirely be described by one dimensionless number, called the inertial number
				$\mathcal{I} = t_{ff}/t_\gamma$.
				This greatly simplifies the possible forms of the constitutive laws, that can be reduced to $\sigma(\mathcal{I})$.
				However, the importance of the third time scale, $t_\Gamma$ is hidden by the Bagnold constraint.
				If we allow other mechanisms for energy injection or dissipation, like a fluidizing force on the granular particles, then
				the Bagnold equilibrium is broken, and another dimensionless number, the Weissenberg number Wi $=t_\Gamma/t_\gamma$ is needed,
				so that the more general form of the laws of rheology is actually $\sigma(\mathcal{I},$Wi$)$.
				This prediction of the GITT model has been successfully tested against experiments in \cite{Angelo25}.

		\end{itemize}

	\subsection{The case of granular suspensions}

		If one has correctly grasped how the rheology is set for dry granular liquids, going to granular suspensions is fairly easy.
		Indeed, the only difference between the two systems concerns what occurs at the microscopic scale between two collision events: is the motion
		of particles that of free falling particles in a pressure field, or that of particles in a viscous medium? In the second case,
		the time-scale for the microscopic motion is $t_\eta = \eta_\infty/P$ (up to some coefficient of order 1 fixed by Darcy's law \cite{Cassar05} that we chose to keep
		equal to 1 as is usual in the literature), where $\eta_\infty$ is the viscosity of the solvent.
		If $t_\eta\ll t_{ff}$, the particles do not really show any big modification of their motion due to the presence of the viscous solvent,
		and the case of dry granular liquids is recovered.
		If $t_\eta\gg t_{ff}$ on the other hand, the microscopic motion of particles is given by the effect of viscous drag, and the particles
		enter the \textit{granular suspension regime}.

		Now, the strength of the analysis performed in the previous subsection is that what happens at the mesoscopic and macroscopic scales is similar in the dry
		and the suspension cases.
		Hence, the rheology is governed by two more time scales which are still $t_\Gamma$ and $t_\gamma$, with the same definitions as in the dry case.
		The different flow regimes are thus:
		\begin{itemize}
			\item The Newtonian regime: $t_\gamma\gg t_\Gamma \simeq t_\eta$.
				In this regime, the constitutive relation takes the form:
				\begin{equation}
					\sigma = \nu_\infty\,\dot\gamma
				\end{equation}
				where $\nu_\infty\propto \eta_\infty$ and is thus strongly influenced by the value of the viscosity of the solvent.
				The prefactor still contains material-dependent coefficient though because the relation $\nu_\infty = t_\Gamma \sigma_y\propto \eta_\infty
				\sigma_y$ holds (the general framework from which the constitutive relations are derived is unchanged, only the value
				of the microscopic time scale is modified).

			\item The yielding regime: $t_\gamma\gg t_\eta$, $t_\Gamma = 0$.
				\begin{equation}
					\sigma \simeq \sigma_y
				\end{equation}
				The behavior of a yielding fluid is recovered. Note that $\sigma_y$ does not explicitly depend on the viscosity of the solvent.

			\item The Bagnold regime: $t_\gamma \ll t_\eta$.
				\begin{equation}
					\sigma = B\,\dot\gamma^2
				\end{equation}

				The derivation of this equation is made exactly in the same way as in the dry granular case.
				There is, however, one major difference between the Bagnold regime in both cases: while a typical experiment on a dry granular
				liquid will automatically take place in the Bagnold regime, in granular suspensions, by definition, the Stokes drag force is
				also present as a source of dissipation, so that the Bagnold equilibrium does not hold, and typical experiments take
				place outside of this regime.

				In terms of dimensional numbers, if we build $\mathcal{J}=t_\eta/t_\gamma$, the constitutive relation now takes the generic form
				$\sigma(\mathcal{J},$Wi$)$.
				Hence, a typical experimental setup for granular suspensions is sensitive to the two dimensionless numbers, in contrast to the
				case of dry granular liquids in which one of them can generally be dropped without problem.
				This is, in our opinion, why attempts to build a unified framework for granular suspensions has failed so far.

				Another remark of importance is that, following the reasoning above, there is only one limit in which the Bagnold regime is observable
				in granular suspensions, that is when the collective effects and advection completely wash out all the traces of the effect of drag
				on the microscopic motion of particles.
				This happens at very high packing fractions, and interestingly, the expression of the shear stress as a function of the shear
				rate matches those of the equivalent dry system \cite{Courrech03,Cassar05}.
		\end{itemize}

		The whole discussion above can be summarized as follows: the constitutive relations for granular suspensions have the same functional form as those of
		dry granular liquids, but with different dependence of the coefficients on the intrinsic properties of the system (typically the viscosity of the solvent).
		Both formulations match in the Bagnold limit, that can be reached in the suspension case only when the collisions are sufficiently numerous so that the
		Stokes drag term in the power balance equation becomes negligible compared to the dissipation by collision term.

\section{The effective Navier-Stokes equation}

	In this section, we are going to explore the consequences of the previously exposed constitutive relations on the modification of the momentum conservation
	equation of the granular fluid, that is build an effective Navier-Stokes equation that captures their complex liquid behavior.

	\subsection{The viscosity tensor in the GITT model}

		The GITT model does not only provide constitutive relations in the form of expressions of the shear stress in terms of the various time scales involved
		in the problem, it allows one to derive the full viscosity tensor $\Lambda_{\alpha\beta\theta\mu}$ linking the stress tensor to the symmetrised
		velocity gradient $\kappa_{\alpha\beta} = \big(\partial_\alpha v_\beta + \partial_\beta v_\alpha\big)$:
		
		\begin{equation}
		\label{eqSigGITT}
			\left<\sigma_{\alpha\beta}\right>^{(\dot\gamma)} = \left<\sigma_{\alpha\beta}\right>^{(0)} + \Lambda_{\alpha\beta\theta\mu}(\dot\gamma)
			\kappa_{\theta\mu}
		\end{equation}
		In this equation, the averages $\left<\cdot\right>$ are ensemble averages that allow to construct quantities at the macroscopic scale,
		the average $\left<\cdot\right>^{(0)}$ is an average taken in an equivalent, fictitious, quiescent (unsheared) liquid at the same temperature
		that the liquid under study -- this last step requires a fluidizing process to balance the dissipation by collisions, for more details about
		this procedure, the reader is referred to the original paper \cite{Kranz20}.

		The first term in Eq.~(\ref{eqSigGITT}) is nothing but the pressure contribution to the stress tensor: $\sigma^{(0)}_{ij} = -P\,\delta_{ij}$.

		Then, the GITT framework is based on the mode-coupling theory \cite{Goetze08}.
		Without entering details that would carry us too far apart from our discussion, let us sum up by saying that the mode-coupling equations
		link the dynamics of system to its internal structure, characterized by two functions: (i) the static structure factor $S_q$, defined from the
		Fourier transform of the particle distribution function $\rho_q$ by $S_q = \left<\rho_q\rho_{-q}\right>$, and (ii) the (normalised) dynamical structure
		factor $\Phi_q(t) = \left<\rho_q(t)\rho_{-q}(0)\right>/S_q$.
		From these two functions, we can built the following combinations:

		\begin{equation}
			\begin{split}
				& \Sigma_\perp = T\big[S_q - S_q^2\big] \\
				& \Delta\Sigma = -T \frac{dS_{q}}{dq}
			\end{split}
		\end{equation}

		Defining the advected wave vector as $q_i(t) = \big(\delta_{ij} + \kappa_{ij}t\big)q_j$,
		the different coefficients of the viscosity tensor are defined as follows:

		\begin{equation}
			\begin{split}
				& \mathcal{B}_1^P = \int_0^{+\infty}dt\int_{\mathbf{q}}\frac{1+\varepsilon}{2S_q^2}\Phi^2_{q(-t)}(t)
				\frac{q^2\Sigma_\perp\Delta\Sigma}{6q(-t)T}t \\
				& \mathcal{B}_2^P = \int_0^{+\infty}dt\int_{\mathbf{q}}\frac{1+\varepsilon}{2S_q^2}\Phi^2_{q(-t)}(t)
				\frac{q^2\Sigma_\perp\Delta\Sigma}{6q(-t)T}t^2 \\
				& \mathcal{B}_X^\sigma = \int_0^{+\infty}dt\int_{\mathbf{q}}\frac{1+\varepsilon}{2S_q^2}\Phi^2_{q(-t)}(t)
				\frac{q^3\Delta\Sigma^2}{30q(-t)T} \\
				& \mathcal{B}_1^\sigma = \int_0^{+\infty}dt\int_{\mathbf{q}}\frac{1+\varepsilon}{2S_q^2}\Phi^2_{q(-t)}(t)
				\frac{q^3\Delta\Sigma^2}{30q(-t)T}t \\
				& \mathcal{B}_2^\sigma = \int_0^{+\infty}dt\int_{\mathbf{q}}\frac{1+\varepsilon}{2S_q^2}\Phi^2_{q(-t)}(t)
				\frac{q^3\Delta\Sigma^2}{30q(-t)T}t^2 \,.
			\end{split}
		\end{equation}

		They are combined in the following way:

		\begin{equation}
		\label{eqTvis}
			\begin{split}
				\Lambda_{\alpha\beta\theta\nu} &= -2\mathcal{B}_1^P\delta_{\alpha\beta}\kappa_{\theta\nu} + \mathcal{B}_2^P\delta_{\alpha\beta}
				\kappa_{\theta i}\kappa_{i \nu} \\
							       &+ \mathcal{B}_X^{\sigma}X_{\alpha\beta\theta\nu} + \mathcal{B}_1^{\sigma}Y^1_{\alpha\beta\theta\nu}
							       +\mathcal{B}_2^{\sigma}Y^2_{\alpha\beta\theta\nu}\,,
			\end{split}
		\end{equation}

		as functions of the tensors built by combining all possible products of Kronecker symbols and symmetrised velocity gradients:

		\begin{equation}
			X_{\alpha\beta\theta\nu} = \delta_{\alpha\beta}\delta_{\theta\nu} + \delta_{\alpha\theta}\delta_{\beta\nu} 
			+ \delta_{\alpha\nu}\delta_{\beta\theta}\,,
		\end{equation}
		and 

		\begin{equation}
			\begin{split}
				& Y^1_{\alpha\beta\theta\nu} = 2\delta_{\alpha\beta}\kappa_{\theta\nu} + \kappa_{\alpha\theta}\delta_{\beta\nu}
				+\kappa_{\beta\nu}\delta_{\alpha\theta} + \kappa_{\alpha\nu}\delta_{\beta\theta} + \kappa_{\beta\theta}\delta_{\alpha\nu} \\
				& Y^2_{\alpha\beta\theta\nu} = \delta_{\alpha\beta}\kappa_{\theta i}\kappa_{i\nu} + \kappa_{\alpha\theta}\kappa_{\beta\nu}
				+\kappa_{\alpha\nu}\kappa_{\beta\theta}\,.
			\end{split}
		\end{equation}

		Expressed in this way, the viscosity tensor is a very complex object to manipulate.
		However, when building effective models, there is one key component that can be used to simplify a lot its use: the wave vector and the time
		are integrated upon.
		Taking the wave number as an example, it means that however complicated the integrand is (there is to the best of our knowledge no good
		analytical approximant of the static structure factor in the liquid state), the $\mathcal{B}_i^\alpha$ coefficients are just constants
		and can be treated as such in an effective equation.
		The question of the time dependence is more delicate since the function $\Phi(t)$ shows a very sharp decay that effectively cuts of the time integral
		at a value that depends on the time scales $t_{ff}$, $t_\eta$, $t_\Gamma$ and $t_\gamma$ (this is actually how they enter the constitutive equations)
		and that depends on the flow regime under consideration.

	\subsection{From GITT to an effective Navier-Stokes equation}

		In the following subsection, we are going to discuss how to generate effective Navier-Stokes equations for non-Newtonian fluids.
		Given that the aim of this paper is not to delve into the technical details of the derivation in order to advocate the GITT model across
		communities, we will pass on the demonstration of some of our claims to reach directly to the results.

		First, it is important to understand that the viscosity tensor given in Eq.~(\ref{eqTvis}) is a fully non-perturbative, non polynomial function of $\dot\gamma$.
		While this could be interesting for some applications like the computation of yield surfaces of granular materials for example \cite{Coquand24},
		it is not the most efficient way to use it in effective models.

		Indeed, if we think of the Navier-Stokes equation as a functional equation of the form
		\begin{equation}
			\partial_t v_\alpha + v_\beta\partial_\beta(v_\alpha) = \partial_\beta\sigma_{\alpha\beta}\big[\mathbf{v}\big]
		\end{equation}
		then the problem can be reduced to finding an expansion of $\sigma\big[\mathbf{v}\big]$ in powers of $\mathbf{v}$ and its derivatives that is
		of low enough order to be as simple as possible, and yet complex enough to allow a description of the whole phenomenology.

		Let us take a concrete example.
		If we cut the expansion of $\sigma$ at the lowest order in powers of $\mathbf{v}$ and its derivatives, then only $\mathcal{B}_X^\sigma$ contributes to
		the viscosity tensor.
		Moreover, we have to treat $\mathcal{B}_X^\sigma$ as a constant, that is, neglect the further dependence on gradients of $\mathbf{v}$ that arise from
		its full expression.
		In this case, the term becomes
		\begin{equation}
			2\mathcal{B}_X^\sigma\, \partial^2v_\alpha = \nu\,\partial^2v_\alpha
		\end{equation}
		that is, we recover the usual Navier-Stokes equation for Newtonian fluids
		(in its incompressible version).

		The further terms in the decomposition will incorporate more and more effects from the fluctuations of the velocity field (that is the deviation
		from the mean flow).
		In order to properly account for the fluctuations, the most appropriate tool is the renormalisation group.
		Again, in order not to get too technical here, we are going to restrict to a sketch of the argument.

		In the renormalisation group framework, once a stable solution of the equations is identified (like the Newtonian solution with Kolmogorov-like
		fluctuations in this case), there is a way to get a hint at whether further terms in the expansion in powers of $\mathbf{v}$ and its derivative
		can drive us away from the stable solution which is called the naive power counting argument.
		The full power counting analysis of the effective Navier-Stokes equation when $\sigma$ is modelled by GITT has been performed in \cite{Coquand25} and
		yields a rather unusual result:
		none of the further terms a priori drive the system away from the Newtonian solution.

		This result is quite surprising since in the Bagnold regime, a different regime of fluctuations has been observed (even though, as emphasized in the introduction,
		more work is required to confirm this scaling law).
		One answer to the apparent paradox is that the power counting argument is not fully rigorous, and non-perturbative effects can lead to violations
		of its conclusions.
		In this case, the next-to-leading order corrections, that appear in the Bagnold regime, should take the following form \cite{Coquand25}:
		\begin{equation}
		\label{eqNSe}
			\begin{split}
				&\partial_tv_\alpha +v_\beta\partial_\beta v_\alpha = - \partial_\alpha(P) + \nu\, \partial^2(v_\alpha) \\
										   &+ \frac{\mathcal{B}^P}{2}
				\Big[\partial_\alpha(\partial_\beta v_\theta)^2 + \partial_\alpha(\partial_\beta (v_\theta)\partial_\theta(v_\beta))\Big] \\
										   &+\frac{\mathcal{B}^\sigma}{2}\Big[\partial_\alpha(v_\theta)\partial^2(v_\theta)
				+ \partial_\theta(v_\alpha)\partial^2(v_\theta)+2\,\partial_\theta(\partial_\beta v_\alpha)\partial_\theta(v_\beta)\Big]
			\end{split}
		\end{equation}
		This equation is a priori the simplest effective model that takes into account the modifications of the constitutive equation in the Bagnold
		regime.
		Note that the Newtonian term is still present, although it is expected that once the Bagnold regime is reached, the strong velocity fluctuations
		will make the following terms dominate over it.

		Finally, let us discuss the yielding regime.
		In this regime, the decay of the two-point density correlation function $\Phi(t)$ is caused by the forced advection of the particles by the
		shear flow, that is, we have $\Phi(t)\simeq\exp\big(-(\dot\gamma t)^2/\gamma_c^2\big)$ \cite{Coquand20g}, where $\gamma_c$ is a typical strain
		scale of the material. Plugging it back into the expressions of the $\mathcal{B}_i^\alpha$ coefficients, we get very roughly:
		\begin{equation}
			\mathcal{B}_X^\sigma \simeq \int_0^{+\infty}dt\,\exp\big(-(\dot\gamma t)^2/\gamma_c^2\big)\times A\simeq A\,\frac{\sqrt{\pi}\gamma_c}{2\dot\gamma}
		\end{equation}
		where $A$ is a constant coming from the $k$ integral.
		A similar procedure can be applied to the other coefficients.

		The conclusion of this analysis is that this time, $\mathcal{B}_X^\sigma\sim\dot\gamma^{-1}$, which is qualitatively very different from the Newtonian case.
		Plugging this back into the expression of the viscosity tensor, we get a dominant contribution in the yielding regime that is independent of $\dot\gamma$
		(or $\kappa$), that is a constant friction term.
		While such a term would contribute only to the computation of the mean flow and not the fluctuations around it, it is very much noteworthy as it
		dominates over all the other terms (unless the velocity fluctuations are so large that the Navier-Stokes basis of the equation does not make sense
		anymore).

		The successive terms in the expansion are then the same as those exhibited for the Bagnold regime.
		Transforming it into an effective model in the yielding regime only requires to add the constant friction term.
		Note that \textit{constant} here means constant with respect to the velocity gradient $\kappa$, local heterogeneities are still allowed, what ensures that
		the derivative of the stress tensor term is a priori non zero.

	\subsection{Application: models of geophysical flows}

		In this subsection, we are going to put the above formalism in application to a concrete case: the modelling of granular flows on the slopes of volcanoes.

		The problem is the following: it has been established that the laws of rheology generally used to describe granular flows in laboratory experiments
		(called frictional or Mohr-Coulomb in the geophysics community, denominations that we will however avoid because they can be ambiguous in the
		context of models in physics).
		A purely phenomenological, but very simple law has been put forward by Dade and Huppert \cite{Dade98} in 1998 whereby the stress term in the Navier-Stokes
		equation is replaced by a constant term.

		Since then, this model has proved to be very useful in various situation. For example, it has been shown to be better suited that the usual laws in
		the modelling of the debris flow on the slopes of the Socompa volcano in Chile \cite{Kelfoun05,Kelfoun08,Davies10}.
		In a slightly different context, it has proven to be also very useful in the modelling of the runout of the pyroclastic flow of the 2006 eruptions
		of the Tungurahua volcano in Ecuador \cite{Kelfoun09}.

		Let us recap what we have established so far.
		Typical experiments on dry granular liquids in laboratory tend to naturally be in the Bagnold flowing regime, a situation that requires that the
		only source of energy injection is shear heating, and the only source of energy dissipation is collision between the granular grains.
		If this particular equilibrium is broken, the system is in the yielding regime if it is has a high enough density, or in a Newtonian regime
		if its density is lower.

		The fact that real geophysical flows are not in the Bagnold regime can be easily explained, particularly in extreme events such as pyroclastic flows:
		triboelectricity, the presence of water, or agitation by very hot gases trapped in the flowing mass of rocks can all provide examples of
		phenomena that can drive the liquid out of the Bagnold regime.
		In this case, the laws of rheology are changed.

		Given the general density of the debris and pyroclastic flows, we expect them to be rather in the yielding than in the Newtonian regime.
		As a further argument in this direction, let us provide a simple explanation of how a model such as that of Dade and Huppert can arise in
		our context.

		Let us consider a mass of geophysical liquid on a cell in a typical depth-averaged model of a flow in a numerical model.
		At the scale of a single cell, we can consider the mass density $\rho$ to be a constant.
		If we call $\mu$ the dynamical friction coefficient of the liquid on the basal layer, and $h(x,y)$ the associated height function,
		the total stress on a cell of surface $S$ is:
		\begin{equation}
			\sigma = \frac{\rho}{S}\,g\mu\,\int_0^x dX\int_0^y dY\, h(X,Y)
		\end{equation}
		so that the terms appearing in the Navier-Stokes equation are:
		\begin{equation}
			\partial_x\sigma = \frac{\rho}{S}\,g\mu\,\int_0^y dY\, h(x,Y)\ \text{and}\ 
			\partial_y\sigma = \frac{\rho}{S}\,g\mu\,\int_0^x dX\, h(X,y)
		\end{equation}
		With the final assumption
		\begin{equation}
			\left|\int_0^y dY\,h(x,Y)\right| \sim \left|\int_0^x dX\,h(X,y)\right|
		\end{equation}
		the Dade and Huppert model is recovered.
		Note that if the last assumption does not hold, we just have to take into account two characteristic stresses in our model, which does not
		increase its complexity.

		As a final note, remark that this model is in full agreement with the predictions of the effective Navier-Stokes equation obtained from the GITT
		model in the yielding regime.
		What this result tells us is that, when such a model is more successful than the Bagnold one in describing a geophysical flow, mechanisms beyond
		shear heating and dissipation by collisions play an important role in the physics of the system.
		Moreover, if the Dade Huppert model is successful, we can also infer that the system is \textit{"rather dense"}, although such a statement is difficult to
		make quantitatively as debris and pyroclastic flows are generally widely polydisperse and the usual orders of magnitude for the packing fraction
		of the granular glass transition surely do not apply here.

\section{Energy repartition in granular suspensions}

	In this last part, we discuss the properties of the velocity fluctuations and related functions outside of the Bagnold regime.

	\subsection{Effective Navier-Stokes equation}

		As far as the effective Navier-Stokes equation is concerned, we are going to discuss only the case of Eq.~(\ref{eqNSe}).
		Indeed, as mentioned before, the constant term arising in the yielding regime is related to the mean value of the velocity field,
		and not to its fluctuations around its mean.
		In the following, it should always be understood that the field $\mathbf{v}$ is the fluctuation field from which the background flow
		has been subtracted.

		We are again going to give a sketch of the arguments without entering too much into the technical details for this work to be accessible to a larger
		audience.
		For more details, the interested reader is referred to \cite{Coquand25}.

		The fluctuations of the velocity field are studied within the formalism of the renormalisation group.
		In this language, each term of the Navier-Stokes equation gets corrections from the fluctuations of the velocity at all the scales smaller than
		the scale at which the system is probed, so that what can be observed in experiments are only effective macroscopic quantities corrected by the
		fluctuations.
		Then, studying carefully the symmetries of the system allows to establish laws that hold at all scales.

		One of the main results of the previous study is that the pressure sector is not renormalised, which means that the pressure term in the Navier-Stokes
		equation is not coupled to the other terms, it does not receive any corrections from the velocity fluctuations, and the velocity is not corrected
		by the pressure fluctuations.

		In the case of granular suspensions, we have established that the viscosity term is replaced by an effective viscosity $\nu_\infty\propto\eta_\infty/P$.
		While, for technical reasons, it is expected that, in the Newtonian regime, $\nu_\infty$ as a whole gets corrected exactly in the same way as $\nu$
		does in the usual Navier-Stokes problem, the interpretation of this result here differs greatly.
		Indeed, $\eta_\infty$ is a property of the solvent, and cannot therefore be corrected by the velocity fluctuations of the
		granular particles.
		Hence, the study of the renormalisation of $\nu_\infty$ tells us how the pressure is corrected by the velocity fluctuations, via a link that
		was absent in the usual Newtonian problem.

		In the limit of the Bagnold regime, we expect the other terms to play a dominant role, which is consistent with the fact that the pictures in
		this limit should match for dry granular liquids and granular suspensions.

	\subsection{The energy cascade picture}

		In Newtonian turbulence, the spectrum of the kinetic energy abides by a solidly established power law $F(k)\sim k^{-5/3}$.
		This power law can be deduced from an argument referred to as the \textit{cascade} picture: vortex structures at large scales give part of their energy to
		structures at smaller scales, until a limit, viscosity dependent, microscopic scale is reached, at which the interaction between the small vortices of 
		different momenta dissipate energy.

		The power law can be derived from the property that the energy dissipation does not depend on the scale at which it takes place \cite{Weizsacker48}.
		Let us reproduce this argument in the case of a granular suspension.

		Let us build a series of scales $l_n$, defined by $l_0=L$, $L$ being of the order of the size of the system, and $l_{n+1}/l_n=\delta$, with $0<\delta<1$.
		We then build the typical velocity of a vortex structure of size $l_n$: $v_n$.
		In an incompressible fluid, the energy dissipated by structures at size $l_n$ is:
		\begin{equation}
			S_n = \eta_{\infty}\big|\mathbf{curl}(\mathbf{v})\big|\sim \eta_\infty v_n^2\,l_n^{-2}
		\end{equation}

		Here lies the crucial difference with the Newtonian case; in ordinary Newtonian fluids, the viscosity entering the formula is scale dependent,
		and its scale dependence, given by Prandtl's Misschungsweghypothese \cite{Prandtl26b}, enters the derivation of the power law.
		Here, the viscosity is fixed to that of the solvent $\eta_\infty$, independent of the scale under consideration.
		If we then enforce that the way energy is dissipated does not depend on scale:
		\begin{equation}
			\frac{dS_n}{dn}=0
		\end{equation}
		we get
		\begin{equation}
			v_n\sim l_n
		\end{equation}

		The kinetic energy $E$ is, by definition, related to its power spectrum $F$ by the following relation:
		\begin{equation}
			E(k) = \int_k^{+\infty}F(k')dk'
		\end{equation}
		All in all,
		\begin{equation}
			\int_k^{+\infty} F(k')dk'\sim k^{-2} \ \Rightarrow\ F(k)\sim k^{-3}
		\end{equation}

		This result is particularly interesting inasmuch as, contrary to the case of the Bagnold regime, the predicted exponent is quantitatively very
		different from the one predicted in the Newtonian case.
		Thus, experiments, or numerical simulations should be able to verify the predictions of our model in a much easier way.

\section{Conclusion}

	To conclude, we have in this work extended the formalism developed previously for the study of dry granular liquids in the Bagnold regime
	to the case of granular suspensions.
	A key to well understand the importance of this problem is that, unlike their dry counterpart, granular suspensions are not, in a typical experimental or natural
	setup, in the Bagnold regime.
	Hence, the Newtonian and yielding regimes, that also exist in dry granular liquids but are scarcely studied, take a larger importance in this problem.

	Our analysis lead us to two key results.
	First, we provided an effective Navier-Stokes equation for granular suspensions in all the flowing regimes that allowed to explain the success of the Dade Huppert
	model in the description of pyroclastic and debris flows (it is expected that dense mud flows obey the same rules).
	Second, we have shown that a direct cascade of kinetic energy is present in granular suspensions in regimes where Stokes drag is the main source of energy dissipation,
	and that the corresponding exponent is qualitatively very different from that of a Newtonian fluid.
	This should ensure that our prediction can be tested easily in experiments or simulations.

\section*{Acknowledgements}

For the purpose of Open Access, a CC-BY 4.0 public copyright licence has been applied by the authors to the present document and will be applied to all subsequent versions up to the Author Accepted Manuscript arising from this submission.

\bibliography{H.bib}

\end{document}